\documentclass[twocolumn,showpacs,preprintnumbers,amsmath,amssymb]{revtex4}

\usepackage{graphicx}
\usepackage{dcolumn}
\usepackage{bm}

\begin{document}

\preprint{APS/123-QED}

\title{Cellular neural networks for NP-hard optimization problems}

\author{M. Ercsey-Ravasz$^{1,2}$, T. Roska$^2$ and  Z. N\'eda$^1$}
\affiliation{$^1$ Babe\c{s}-Bolyai University, Faculty of Physics, RO-400084, Cluj, Romania \\
     $^2$ P\'eter P\'azm\'any Catholic University, Faculty of Information Technology, HU-1083, Budapest, Hungary }

\date{\today}

\begin{abstract}
Nowadays, Cellular Neural Networks (CNN) are practically 
implemented in parallel, analog computers, showing a fast developing trend. Physicist must be aware that 
such computers are appropriate for solving in an elegant manner practically important problems, which are 
extremely slow on the classical  digital architecture.  Here, CNN is used for solving NP-hard optimization 
problems on lattices. It is proved, that a CNN in which the parameters of all cells can be separately controlled, 
is the analog correspondent of a two-dimensional Ising type (Edwards-Anderson) spin-glass system. Using the 
properties of CNN computers a fast optimization method can be built for such problems. Estimating the simulation 
time needed for solving such NP-hard optimization problems on CNN based computers, and comparing it with the 
time needed on normal digital computers using the simulated annealing algorithm, the results are 
astonishing: CNN computers would be faster than digital computers already at $10\times 10$ lattice sizes. 
Hardwares realized nowadays are of $176\times 144$ size. Also, there seems to be no technical difficulties 
adapting CNN chips for such problems and the needed local control is expected to be fully developed in 
the near future.
\end{abstract}

\maketitle

\section{\label{sec:level1} Introduction}

Solving NP-hard problems is a key task when testing novel computing paradigms. 
These complex problems frequently appear in physics, life sciences, biometrics, logistics, database search...etc., 
and in most cases they are associated with important practical applications as well \cite{NP}. The deficiency of 
solving these problems in a reasonable amount of time is one of the most important limitation of digital computers, 
thus all novel paradigms are tested in this sense. Quantum computing for example is theoretically well-suited 
for solving NP-hard problems, but the technical realization of quantum  computers seems to be quite hard.
Here we present a computing paradigm based on Cellular Neural Networks (CNN) with good perspectives 
for fast NP-hard optimization. The advantage of this approach relative to quantum computing
is that several practical realizations are already available. 

After the idea of CNN appeared in 1988 
\cite{chua:cnn88} a detailed plan for a CNN computer was developed in 1993 \cite{roska:cnnum93}. 
Since than the chip had a fast developing trend and the latest - already commercialized - 
version is the EYE-RIS chip with lattice size $176\times 144$, mainly used as a visual microprocessor \cite{Eye-Ris}.
The physics community can also benefit from CNN based computers. 
It has been proven in several previous studies that this novel computational paradigm is 
usefull in solving partial differential equations \cite{pdeI,pdeII}, studying cellular automata models \cite{popdyn, crounse:rand96}, doing 
image processing \cite{applications} and making Monte Carlo simulations on lattice models \cite{cnnperc,cnnising}. 
Here it is shown that NP-hard optimization problems can be also effectively solved using such an approach.  

In the next section we will present briefly the structure and dynamics of cellular neural networks together with  
the most developed applications of CNN computing. In the third section we prove that a CNN computer 
on which the parameters of each cell can be separately controlled, is the analog correspondent of 
a locally coupled two-dimensional spin-glass system. Using the properties of CNN computers a fast optimization 
algorithm can be thus developed. The local control of the parameters of each cell is already partially 
realized on some hardwares (for details see section 2.) and it is expected to be fully functional in the near future. 
Beside the fundamental interest for solving NP hard problems, the importance of this study 
consists also in motivating the development of hardwares in such direction.

\section{\label{sec:level2} Cellular Neural Networks and CNN Computers}

The CNN Universal Machine (CNN-UM) \cite{roska:cnnum93} is one special case of cellular wave computers 
\cite{roska:cellwavecomp}  in which computation is achieved using the spatial-temporal dynamics of  
a  cellular neural network  \cite{chua:cnn88}. The CNN  is composed by $L \times L$ cells 
placed on a square lattice and interconnected through their neighbors \cite{chua:cnn88}. Usually the $4$ nearest and the 
$4$ next-nearest neighbors (Moore neighborhood) are considered. Each cell is characterized by a 
state value: $x_{i,j}(t)$ 
representing a voltage in the circuit of the cell. The cell has also an input value (voltage) $u_{i,j}$, 
which is constant in time and can be defined at the beginning of an operation. The third characteristic 
quantity of the cell is the output value $y_{i,j}(t)$. This is equivalent with the $x_{i,j}$ state value 
in a given range. More specifically it is a piece-wise linear function, bounded between $-1$ (called as white) 
and $1$ (black): $y=f(x)\equiv \frac{1}{2} ( \mid x+1 \mid - \mid x-1 \mid )$.

The wiring between neighboring cells assures that the state value of each cell can be influenced by the 
input and output values of its neighbors. The equation governing the dynamics of the CNN cells 
results from the time-evolution of the equivalent circuits. Supposing the $8$ Moore neighbor interactions 
it has the following form \cite{chua:cnn88}:
\begin{eqnarray}
\frac{d x_{i,j}(t)}{d t}=-x_{i,j}( t ) + \sum_{k=i-1}^{i+1} \sum_{l=j-1}^{j+1} A_{i,j;k,l} y_{k,l} ( t )+ \nonumber \\ 
+\sum_{k=i-1}^{i+1} \sum_{l=j-1}^{j+1} B_{i,j;k,l} u_{k,l}  + z_{i,j}  \label{state}
\end{eqnarray}
where ${i,j}$ denotes the coordinates of the cell and the summation indeces ${k,l}$ are for its neighbors. 
Self-interaction ($k=i,l=j$) is also possible. The set of parameters $\{A,B,z\}$ is called a template and 
controls the whole system. An operation is performed by giving the initial states of the cells, the input image 
(the input values of all cells) and by defining a template. The states of all cells will vary in parallel 
and the result of the operation will be the final steady state of the CNN. If the state values ($x_{i,j}$) 
of all cells remain bounded in the $[-1,1]$ region  (i.e. $y_{i,j}=x_{i,j}$ holds for each cell at any time $t$), 
than each operation is equivalent with solving a differential equation defined by the template 
itself \cite{chuaroskakonyv,pdeI,pdeII}. When $x_{i,j}$ does not remain bounded, than the piece-wise linear function 
described at the definition of the output value $y_{i,j}$, takes an important role. The final steady state will 
not be simply the solution of the differential equation, and this case can be used for defining other 
useful operations as well \cite{chuaroskakonyv}.

The CNN-UM \cite{roska:cnnum93} is a programmable analogic (analog \& logic) cellular wave computer. 
Beside the analog circuits described by Eq.\ref{state}, each cell contains also a logic unit, local analog and 
logic memories and  a local communication and control unit. The logic unit and logic memories are  included 
to complement the analog computation. In this manner basic logic operations can be performed without 
defining complicated templates for it. In the local logic memories one can save a binary value 
($1$ and $0$ respectively), and in the local analog memories it is possible to save real values between $-1$ and $1$. 
Since the CNN array is mainly used for image processing and acquisition, the binary values are often 
referred as black and white, and the real values bounded between $-1$ and $1$ are  mapped in a gray-scale scheme.    
Beside these local units, the CNN-UM has also a global analog programming unit which controls the whole system, 
making it a programmable computer. It can be easily connected to PC type computers and programmed with special 
languages, for example the Analogic Macro Code (AMC).

The physical implementations of these computers are numerous and widely different: mixed-mode CMOS, emulated 
digital CMOS, FPGA, and also optical. For practical purposes the most promising applications are for image 
processing, robotics or sensory computing purposes \cite{applications}, so the main practical drive in the mixed-mode implementations 
was to build a visual microprocessor \cite{chuaroskakonyv}.  In the last decades the size of the engineered 
chips was constantly growing, the  new cellular visual microprocessor EYE-RIS \cite{Eye-Ris} for example has 
$176\times 144$  processors, each cell hosting also $4$ optical sensors. Parallel with increasing the lattice 
size of the chips, engineers are focusing on developing multi-layered, $3$ dimensional chips as well.
 
For the first experimental versions the templates (coupling parameters) are defined identical for all cells. 
This means that for example $A(i,j;i+1,j)$ is the same for all $(i,j)$ coordinates. In such way, 
on the two-dimensional CNN chip, all the $A$ couplings are defined by a single $3\times 3$ matrix. Totally, 
$9+9+1=19$ parameters are needed to define the whole, globally valid, template (\{A,B,z\}). On the latest 
version of the CNN chips  (ACE16K, EYE-RIS) the $z(i,j)$ parameter can already be locally varied. It is  
expected that on newer chips one will be able to separately control also the $A(i,j;k,l)$ and 
$B(i,j;k,l)$ connections as well. 

Many applications ideal for the analogic and parallel architecture of the CNN-UM were already developed and tested. 
Studies dealing with  partial differential equations  \cite{pdeI,pdeII} or cellular automata  models 
\cite{popdyn,crounse:rand96} prove this. In some of our latest publications we have shown that CNN computers 
are suitable also for stochastic simulations. The natural noise of the analog chip can be effectively 
used to generate random numbers approximately $4$ times faster than on digital computers \cite{cnnperc}. 
We also presented experiments in which the site-percolation problem and the two-dimensional Ising model 
was properly solved on the ACE16K chip (with $128\times 128$ cells) \cite{cnnperc,cnnising}.  

\section{\label{sec:level3} NP-hard optimization on CNN architectures}

The aim of the present study is to prove that CNN computing is also suitable for solving 
effectively complex optimization problems on spin-glass type lattice models. We consider a two-dimensional 
CNN where the templates (parameters of Eq. \ref{state}) can be locally controlled. Matrix $A$ is considered symmetric
$A(i,j;k,l)=A(k,l;i,j)$, $A(i,j;i,j)=1$ for all $(i,j)$, and the elements are bounded
$A(i,j;k,l) \in [-1,1]$  ($(i,j)$ and $(k,l)$ denote two neighboring cells). Matrix $B$, which controls  
the effect of the input image, will be taken simply as: $B(i,j;i,j)=b$ and 
$B(i,j;k,l)=0$, $\{i,j\}\neq \{k,l\}$. The parameter $z$ is chosen as $z=0$, so finally our template 
is defined by $\{A,b \}$ alone. 

In an earlier work Chua {\it et al.} \cite{chua:cnn88} defined a Lyapunov function for the CNN, which behaves 
like the "energy" (Hamiltonian) of the system. For the CNN defined above it can be written simply as
\begin{equation}
E(t)=-\sum_{<i,j;k,l>} A_{i,j;k,l} y_{i,j} y_{k,l}-b \sum_{i,j} y_{i,j}u_{i,j},
\label{energy}
\end{equation}
where $<i,j;k,l>$ denotes pairs of Moore neighbors, each pair taken only once in the sum. $y_{i,j}$ denotes the
output value of each cell  and $u_{i,j}$ stands for an arbitrary input image. By choosing the 
parameter $b=0$, the energy of this special CNN is similar with the energy of an Ising type 
system on square lattice with locally varying coupling constants. The difference 
is that Ising spins are defined as $\pm 1$, while here we have continuous values between $[-1,1]$. Since 
the $A(i,j;k,l)$ coupling constants can be positive and negative as well, locally coupled spin-glasses 
can be mapped in such systems. In the following we will be especially interested in the case when the 
$A(i,j;k,l)$ couplings lead to a frustration and the quenched disorder in the system is similar with that of 
spin-glass systems (\cite{edwand,sherkirk}).  
 
The Lyapunov function defined by Chua {\it et al.} has two important properties \cite{chua:cnn88}: 1.)  
it is always a monotone decreasing function in time, $dE/dt\le 0$, so starting from an initial condition $E$ can 
only decrease during the dynamics of the CNN. 2.) the final steady state is a local minimum of the energy: $dE/dt=0$.
In addition to these, our CNN has also another important property: due to the fact that all self-interaction 
parameters are $A(i,j;i,j)=1$,  it can be shown that the output values of the cells in a final steady state will 
be always either $1$ or $-1$. The local minima achieved by the 
CNN is thus an Ising-like configuration. We can conclude thus that starting from any initial condition the 
final steady state of the template - meaning the result of an operation - 
will be always a local minimum of the spin-glass type Ising spin system with local connections 
defined by matrix $A$.  The fact that one single operation is needed for finding a local minimum of the energy, 
gives us hope to develope fast optimization algorithms.

As already emphasized, the complex frustrated case (locally coupled spin-glass type system), where the $A$ coupling 
parameters generates a non-trivial quenched disorder, will be considered here. The
minimum energy configuration of such systems is searched by an algorithm which is similar with the
well-known simulated annealing method \cite{simaneal}. The noise is included with random input images 
($u_{i,j}$ values in eq. \ref{energy}) acting as an external locally variable magnetic field. The strength of this
field is governed through parameter $b$. Whenever $b$ is different from zero, our CNN template minimizes 
the energy with form \ref{energy}: the first part of it being the energy of the considered spin-glass type model 
and the second part an additional term, which gets minimal when the state of the system is equal to the input 
image (the external magnetic field). If $b$ is large, the result will be the input image itself, 
if $b=0$ the result is a local minimum of the pure Ising-type system. For values in between, our result is 
a "compromise" between the two  cases. Slowly decreasing the value of $b$ will result in a process similar with 
simulated annealing, where the temperature of the system is consecutively lowered. First big fluctuations 
of the energy are allowed, and by decreasing this we slowly drive the system 
to a low energy state. Since the method is a stochastic one, we can of course never be totally sure that the global 
minimum will be achieved. 

The steps of the algorithm are the following: \\
 1. One starts from a random initial condition $x$, and $b=5$ (with this value the result of the template 
 is almost exactly the same as the input image).\\
 2. A binary random input image $u$ is generated with $1/2$ probability of black ($1$) pixels, \\
 3. Using the $x$ initial state and the $u$ input image the CNN template is applied, \\
 4. The value of $b$ is decreased with steps $\Delta b$, \\
 5. Steps 2-4 are repeated until $b=0$ is reached. The results of the previous step (minimization) is considered 
    always as the initial state for the next step.\\
 6. When reaching $b=0$ the image (Ising spin configuration) is saved and the energy is calculated.

In the classical simulated annealing algorithm several thousands of steps for a single temperature are needed. 
Here the CNN template working totally in parallel replaces all these steps. Similarly with choosing the cooling rate 
in simulated annealing, choosing the value of $\Delta b$ is also a delicate problem. A proper value  
providing an acceptable compromise between the quality of the results and speed of the algorithm has to be found. 
For each system size one can find an optimal value of $\Delta b$, but as one would expect this is 
rapidly decreasing by increasing the system size. It is much more effective, both for performance 
(meaning the probability of finding the real global optimum) and speed, to choose a constant 
$\Delta b=0.05$ step and repeat the whole cooling process several times. As a result, several different final states 
will be obtained, and we have a higher probability to get the right global minima between these.  

For testing the efficiency of the algorithm one needs to measure the number of steps necessary for finding 
the right global minima. To do this, one has to previously know the global minima. In case of small systems 
with $L=5,6$ this can be obtained by a quick exhaustive search in the phase-space. For bigger systems the 
classical simulated annealing algorithm was used. The temperature was decreased with a rate of $0.99$ 
($T_{final}/T_{ini}$) and $1000$ Monte Carlo steps were performed for each temperature.
  
\begin{figure}                                  
\includegraphics[width=0.5\textwidth]{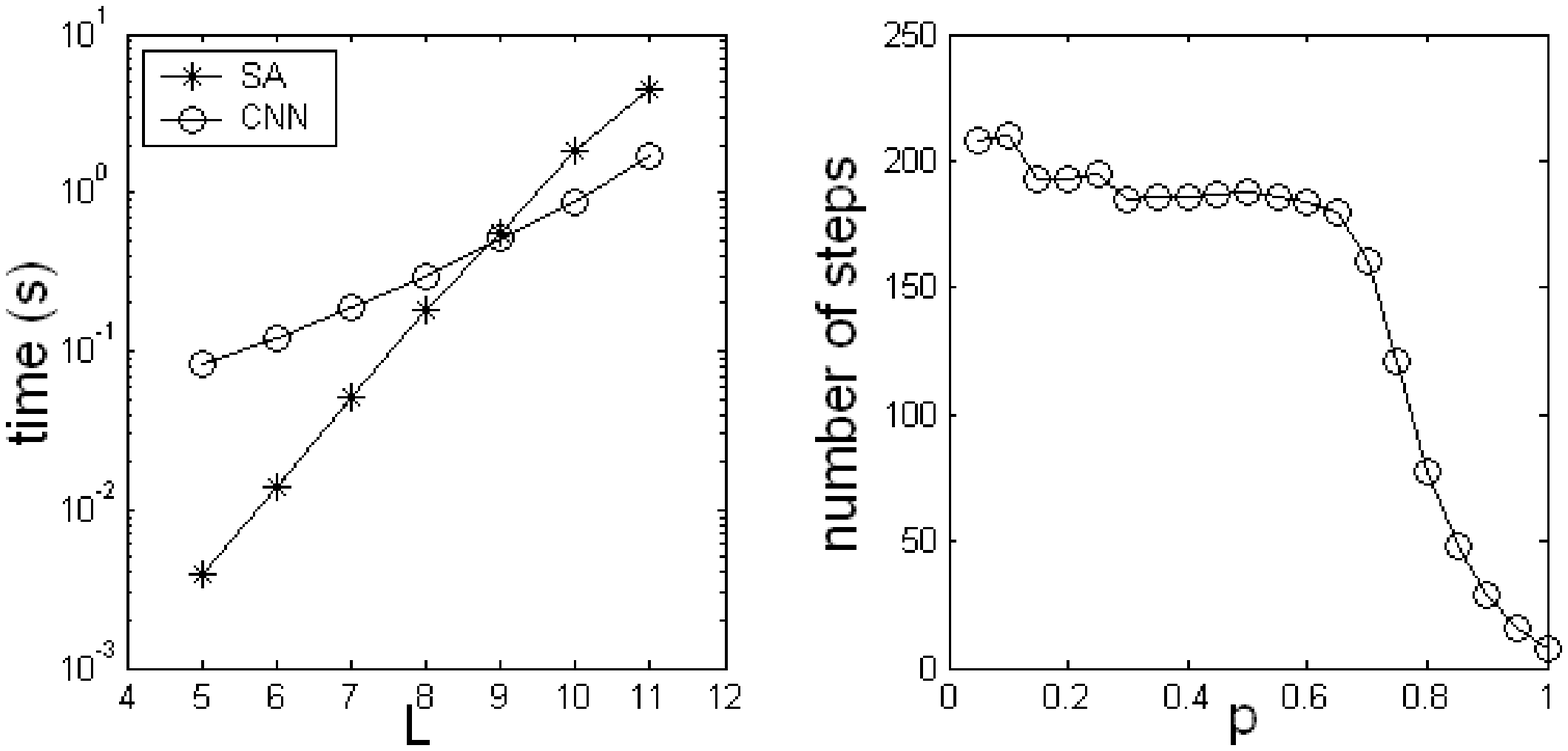}
\caption{a) Time needed to reach the minimum energy as a function of the lattice size $L$. 
Circles are for estimates on CNN computers and stars are simulated annealing results on 3.4 GHz Pentium 4.
Results are averaged on $10000$ different configurations with $p=0.4$ probability of positive bonds.
For the CNN algorithm $\Delta b=0.05$ was chosen. For simulated annealing the initial temperature was 
$T_0=0.9$ , final temperature $T_f=0.2$ and the decreasing rate of the temperature was fixed as $0.99$.
b) Number of steps needed for reaching the presumed global minima 
as a function of the probability $p$, of positive connections. 
Simulation results with the proposed algorithm on a CNN lattice with size $L=7$. }
\label{L-p}
\end{figure}

In the present work spin-glass systems with $A(i,j;k,l)=\pm 1$ local connections were studied. The $p$ probability 
of the positive bonds was varied (influencing the amount of frustration in the system), and local interactions 
with the $8$ Moore neighbors were considered. For several $p$ densities of  
the positive links and various system sizes, we calculated the average number of steps needed for 
finding the energy minimum. As naturally is expected for the non-trivial frustrated cases, the needed simulation 
time exponentially increases with the system size. As an example, the $p=0.4$ case is shown 
on Fig. \ref{L-p}a. Circles represent the estimated time 
on CNN computers and stars illustrate the time measured on a Pentium 4, 3.4 GHz computer using 
the classical simulated annealing method. For the calculation time on CNN computers we have used a 
convention (arguments given in the next paragraph) that roughly $1000$ simulation steps are made in $1$ second.  
As observable in the figure, we could made estimates for relatively small system sizes only ($L\ge 12$). 
The reason for this is that we had to simulate also the operations on the CNN chip, and for 
large lattices a huge system of partial differential equations had to be solved. This process gets 
quite slow for bigger lattices. 

The needed average number of steps to reach the estimated energy minima depends also on the $p$ probability 
of the positive connections in the system. On Fig.\ref{L-p}b we illustrate this for a system with size $L=7$.
To obtain this data for each $p$ value $5000$ different systems were analyzed. 
As observable on Fig.\ref{L-p}b the system is almost equally hard to solve for all $p$ values 
in the rage of $p \in (0,0.6)$.

Finally, let us have some thoughts about the estimated speed of such an optimization algorithm. 
As mentioned earlier, on the nowadays available CNN chips, only parameter $z$ can be locally varied, 
parameters $A$ and $B$ are $3\times 3$ matrices, uniformly applied for all cells. The reason for no local control 
of $A$ and $B$ seems to be simply the lack of motivations. In image processing applications 
no really useful algorithms were developed, which would require these locally variable connections. 
Realizing the local control of $A$ and $B$ is technically possible and is expected to be 
included in the newer versions of the CNN chips. This modification would not change the properties 
and the speed of the chip, only the control unit and template memories would become more complicated. 
Also, introducing the connection parameters in the local memories of the chip would 
take a slightly longer time. In the specific problem considered here the connection parameters have 
to be introduced only once for each problem, so this would not effect in a detectable manner the speed of 
calculations. Based on our previous experience with the ACE16K chip (with sizes $128 \times 128$) 
\cite{cnnperc,cnnising} we can make an estimation of the speed for the presented optimization algorithm. 
This chip with its parallel architecture solves one template in a very short  
time - of the order of microseconds. For each step in the algorithm one also needs to generate 
a random binary image. This process is already $4$ times faster on the ACE16K chip 
than on a 3.4 GHz Pentium 4 computer and needs around $100 \mu s$ (see \cite{cnnperc}). 
It is also helpful for the speed, that in the present algorithm it is not needed to save information at each step, 
only once at the end of each cooling process. Saving an image takes roughly $10$ milliseconds on the ACE16K, 
but this is done only once after several thousand of simulation steps. 
Making thus a rough estimate for our algorithm, a chip with similar properties like the  ACE16K should be able to compute between 
$1000-5000$ steps in one second, \emph{independently of the lattice size}.
Using the lower estimation value ($1000$ steps /second) and following up the number of steps needed 
in case of $p=0.4$, the estimated average time for solving one problem is plotted as a function of the lattice 
size  in Fig. \ref{L-p}a (empty circles). Comparing this with the speed of simulated 
annealing (SA) performed on a $3.4$ GHz Pentium 4 (stars on Fig. \ref{L-p}a), the results for larger lattice sizes 
are clearly in favour of the CNN chips. For testing the speed of simulated annealing we used 
the following parameters: initial temperature $T_0=0.9$ , final temperature $T_f=0.2$,  
decreasing rate of the temperature $0.99$. Results were avereged for $10000$ different bond distributions. 
From Fig. \ref{L-p}a it results that the estimated time needed for the presented algorthim on a CNN 
chip would be smaller than simulated annealing already at $10\times 10$ lattice sizes. 
 
Spin-glass like systems have many applications in which global minimum is not crucial to be exactly found, 
the minimization is needed only with a margin of error. In such cases the number of requested steps will 
decrease drastically. As an example in such sense, it has been shown that using spin-glass models 
as error-correcting codes, their cost-performance is excellent \cite{sourlas}, and the systems 
are usually not even in the spin-glass phase. In this manner by using the CNN chip, finding acceptable 
results could be very fast, even on big lattices.

\section{Conclusion}
A cellular neural network with locally variable parameters was used for finding the optimal state of locally 
coupled, two-dimensional, Ising type spin-glass systems. By simulating the proposed optimization algorithm on
a CNN chip, where all connections can be locally controlled, very good perspectives for solving such 
NP hard problems were predicted: 
CNN computers could be faster than digital computers already at a $10\times 10$  lattice size. 
Chips with $2$ and $3$ layers of cells were also produced (CACE1k, XENON) and increasing the number of 
layers is expected in the near future. This further extends the number of possible applications. 
On two layers is possible to map already a spin system with any connection matrix (even globally coupled spins) 
and also other important NP-hard problems (e.g. K-SAT) may become treatable.

\begin{acknowledgments}
Work supported from a Romanian CNCSIS No.1571 research  grant (contract 84/2007) 
and a Hungarian ONR grant (N00014-07-1-0350).
\end{acknowledgments}

\end{document}